\RequirePackage{tikz}
\documentclass[sn-mathphys]{sn-jnl}% Math and Physical Sciences Reference Style
\usepackage{subcaption}
\jyear{2021}%

\theoremstyle{thmstyleone}%
\theoremstyle{thmstyletwo}%

\theoremstyle{thmstylethree}%

\begin{document}

\title[Enhanced drug uptake on application of electroporation]{Enhanced drug uptake on application of electroporation in a single-cell model}


\author[1]{\fnm{Nilay} \sur{Mondal}}\email{nilay.mondal@iitg.ac.in}
\equalcont{These authors contributed equally to this work.}

\author*[1]{\fnm{K. S.} \sur{Yadav}}\email{yadav176123004@iitg.ac.in}
\equalcont{These authors contributed equally to this work.}

\author[1]{\fnm{D. C.} \sur{Dalal}}\email{durga@iitg.ac.in}
\equalcont{These authors contributed equally to this work.}


\affil*[1]{\orgdiv{{Department of Mathematics}, \orgname{Indian Institute of Technology Guwahati}, \orgaddress{\street{Amingaon}, \city{Guwahati}, \postcode{781039}, \state{Assam}, \country{India}}}}

%\affil[2]{\orgdiv{Department}, \orgname{Organization}, \orgaddress{\street{Street}, \city{City}, \postcode{10587}, \state{State}, \country{Country}}}
%
%\affil[3]{\orgdiv{Department}, \orgname{Organization}, \orgaddress{\street{Street}, \city{City}, \postcode{610101}, \state{State}, \country{Country}}}


\abstract{Electroporation method is a useful tool for delivering drugs into various diseased tissues in the human body. As a result of an applied electric field, drug particles enter the intracellular compartment through the temporarily permeabilized cell membrane. Consequently, electroporation method allows better penetration of the drug into the diseased tissue and improves treatment clinically. In this study, a more generalized model of drug transport in a single-cell is proposed. The model is able to capture non-homogeneous drug transport in the cell due to non-uniform cell membrane permeabilization. Several numerical experiments are conducted to understand the effects of electric field and drug permeability on drug uptake into the cell. Through investigation, the appropriate electric field and drug permeability are identified that lead to sufficient drug uptake into the cell. This model can be used by experimentalists to get information prior to conduct any experiment, and it may help reduce the number of actual experiments that might be conducted otherwise.}




\maketitle

\section{Introduction}
In order to cure a disease, one aims to deliver a sufficient amount of drug to the diseased site and into the infected cells. The cell membrane is selective permeable in the sense that it does not allow all the molecules to pass through it. The permeability of the drug across the cell membrane depends on the properties of the drug as well as on the membrane pores.

Over the years, several methods have been developed for drug delivery. In the advancement of technology, new techniques, such as electroporation, micro-injection, laser, ultrasound are developed \cite{Bolhassani2011}. Electroporation has been used widely in \textit{in-vitro} and \textit{in-vivo} models in various applications \cite{Miklavcic2018Feb,Kotnik2019,Miklavcic2020}.  In electroporation, cell membrane is temporarily destabilized by the application of external electric field. This destabilization occurs due to increment in the transmembrane potential \cite{Neumann}. In the destablization process, the nano-meter size pores are created in lipid bilayers \cite{Krassowska2007,Miklavcic2014,Miklavcic2016,Mondal2021}. The transitory and permeabilized states of the cell membrane can be exploited to transport drugs into the intracellular domain. 

The physical properties of the tissue, such as cell shape, size and its distribution, as well as the electrical parameters, such as the number of pulses, pulse amplitude,  pulse duration and tissue conductivity, influence drug transport into the targeted cells \cite{Pavlin,Pucihar2011}. Cell electroporation is commonly used with short and long duration pulses in various experiments. To avoid cell death, single-cell electroporation that involves an inhomogeneous short-duration low-voltage electric field around the cell surface is generally applied. However, the application of high voltage electric field enhances cell membrane pore density \cite{Pavlin,Weaver2003}. 

For drug absorption in electroporated cells, it is vital to focus on the increase in permeability and control of cell death. 
%Researchers are continuously working to improve mathematical models for these components. 
Granot and Rubinsky \cite{Granot}  proposed a model to investigate drug delivery in a tissue  with electroporation. In their study, a mass transfer coefficient in terms of pore density based on the pore creation model \cite{Krassowska2007} has been developed. Goldberg et al. \cite{Goldberg2018} proposed a multiphysics model for ion transport, which is also based on the pore creation model.	They presented a mass transport model using the Nernst-Planck equation for transporting various species into cells in their model. Goldberg et al. \cite{Goldberg2021} extended their previous model and described the effects of electric pulses on cisplatin transport across the plasma membrane.  The model shows that an electric field induces maximum transmembrane potential  at the cell poles where the electrodes are placed. 

It is a challenging and difficult task to conduct experimental studies of pulse application and field strength on single cell to improve drug uptake. The detailed investigation to determine appropriate class of drugs that can easily enter the cell in a desired amount is still missing in the literature. 

%It is important to improve the mathematical model in this area of research for application in cancer and tumour treatment.

In the present article, a mathematical model is employed to study drug transport into a single-cell by the application of electroporation. The prescribed model is more generalized  in comparison to the previous models as the spatial changes in cellular drug uptake at different times are analyzed here. 
%A general model is presented in comparison to the literature, where a simplified model in spherical coordinates was proposed. 
This study presents a rigorous analysis of drug delivery into a diseased cell emphasizing  the effects of membrane resealing. Drug transport takes place due to diffusion from the extracellular space to the intracellular one. Transport across the cell membrane takes place due to passive diffusion, which means that the diffusion takes place due to the concentration gradient and the ability of the cell membrane (permeability) to pass the drug. 
%The model is solved on Cartesian coordinates incorporating a spherical cell inside the computational domain and it provides an advanced solution method for these types of models. 
The drug transport equations are solved using the permeable interface method (PIM) on Cartesian mesh by treating the cell membrane as a sharp interface. Various electrode arrangements are tested to determine the locations of maximum pore formation and to analyze the mass  transportation through those locations.  The numerical experiments are conducted with various electric fields and multiple pulses for a given electric field are used in electroporation to reduce the resealing effects on drug uptake. From the investigation, a suitable electric field is determined to improve the cellular uptake for a specific drug. The numerical experiments are also conducted with drugs of different permeabilities to select the appropriate drug that can be used for treatment to get the best possible efficacy. The resulted non-uniformity of the pores in the cell membrane owing to the application of electric pulses leads to non-uniform drug distribution in the intracellular space. 

\section{Model formulation}
\begin{figure}[h!]
	\centering
	\begin{tikzpicture}[scale=0.5]
		
		\draw[fill,blue!30] (-6,-6) rectangle (6,6);
		\draw[ultra thick] (-6,-6) rectangle (6,6);
		
		\draw[fill,yellow!40](0,0) circle (2cm);
		\draw[ultra thick] (0,0) circle (2cm);
		\draw[ultra thick] (0,0) circle (2.1cm);
		
		\draw[ultra thick, ->] (-10,4) -- (-6,4);
		\draw[ultra thick, ->] (-10,0) -- (-6,0);
		\draw[ultra thick, ->] (-10,-4) -- (-6,-4);
		
		\draw (-8,2) node{\large Drug};
		\draw (0,0) node{\large Cell};
		\draw[ultra thick, ->] (0,-3) -- (0,-2.1);
		\draw (0,-3.4) node{\large Cell Membrane};
		\draw (0,3) node{\large Extracellular Space};
		\draw (5,-5) node{\large $\Omega$};
		
	\end{tikzpicture}
	
	\caption{A schematic diagram of injecting drug  into a biological tissue.}
	\label{tissue1}	
\end{figure}
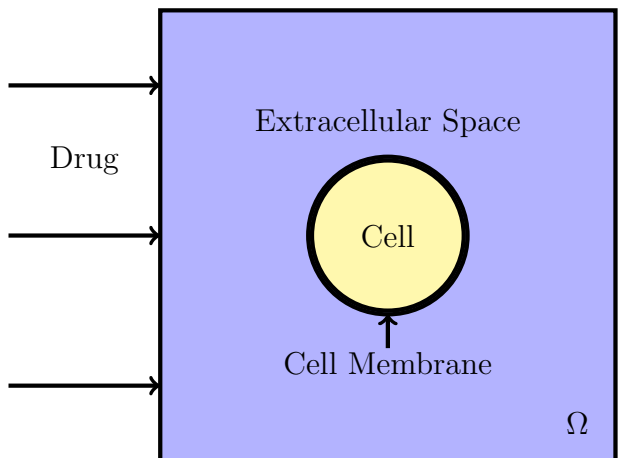

This study investigates the drug transport through the electro-permeabilized cell membrane. A square domain ($\Omega$) of edge length $L$ is considered, and a single-cell is assumed to be placed at the center of $\Omega$. Structurally, the domain can be viewed as two parts: extracellular space and intracellular space. The spaces are separated by the cell membrane, which is selective permeable and controls the mass exchange between the extracellular and intracellular domains. The schematic diagram is shown in the Fig. \ref{tissue1}. 

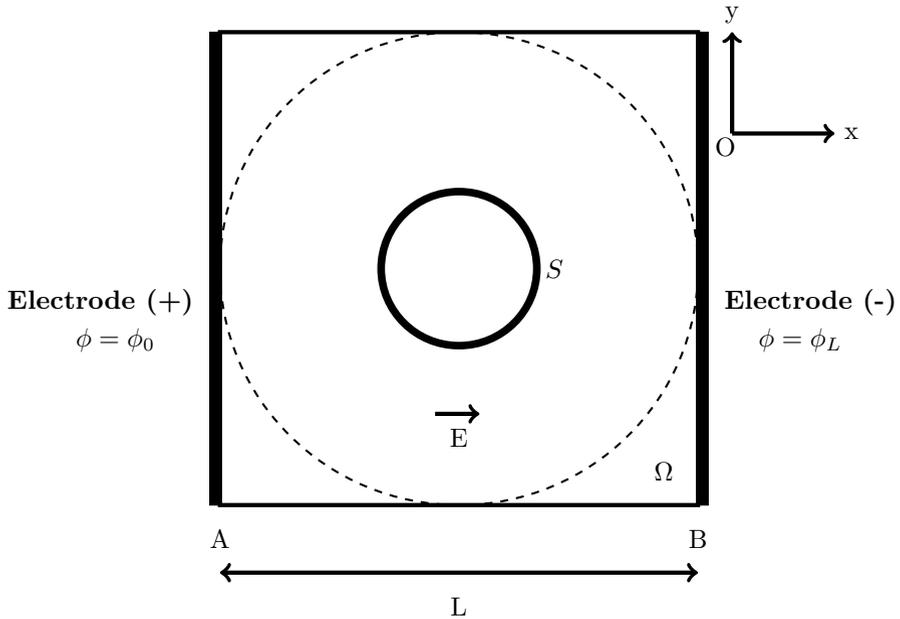
\begin{figure}[h!]
	\centering
	\begin{tikzpicture}[scale=0.45]
		
		\draw[fill,black!100] (-6.3,-6) rectangle (-6,8);
		\draw[fill,black!100] (8,-6) rectangle (8.3,8);
		\draw[ultra thick] (-6,-6) rectangle (8,8);
		
		\draw[ultra thick] (1,1) circle (2.23cm);
		\draw[ultra thick] (1,1) circle (2.33cm);
		\draw[thick,dashed] (1,1) circle (7cm);

		%	\draw[fill,black] (-2,0) circle (0.02cm);
		%	\draw[] (-2,0) -- (0,0);
		%	\draw[] (-2,0) -- (-1,1.7);
		%	\draw[] (-1.4,0) .. controls (-1.3,0.35) .. (-1.7,0.5);
		%	\draw (-0.9,0.5) node[]{$\Psi$};
		%	\draw (-2,-3) node[]{Cell};
		
		\draw (-6,-7) node[]{A};
		\draw (8,-7) node[]{B};
		
		\draw[ultra thick,<->](-6,-8) -- (8,-8);
		\draw (1,-9) node[]{L};
		%\draw[ultra thick,<->](12,-8) -- (12,8);
		%\draw (-11,3) node[]{\large H};
		\draw[ultra thick,->](0.3,-3.3) -- (1.6,-3.3);
		\draw (1,-4) node[]{E};

		\draw (11.3,0) node[]{\bf Electrode (-)};
		%\draw (1.2,-2) node[anchor=north west]{\large $r=r_0$};
		\draw (9.5,-0.5) node[anchor=north west]{$\phi=\phi_L$};
		
		\draw (-9.5,0) node[]{\bf Electrode (+)};
		%\draw (-12,-2) node[anchor=north west]{\large $r=R$};
		\draw (-10.5,-0.5) node[anchor=north west]{$\phi=\phi_0$};
		
		\draw[ultra thick,->](9,5) -- (9,8);
		\draw[ultra thick,->](9,5) -- (12,5);
		\draw (12.5,5) node[]{x};
		\draw (9,8.5) node[]{y};
		\draw (8.8,4.6) node[]{O};
		
		\draw (7,-5) node[]{$\Omega$};
		\draw (3.8,1) node[]{$S$};
		
	\end{tikzpicture}
	
	\caption{Schematic representation of a single cell electroporation.}
	\label{electrode1}
\end{figure}

In order to electroporate the cell membrane, two electrodes with potential values $\phi_0$ and $\phi_L$ are placed along the vertical lines at A and B, respectively (as shown in Fig. \ref{electrode1}). A uniform electric field $E$ is induced in the region directed from positive to negative electrode. The transmembrane potential ($V_m$) increases due to the induced electric field. On increasing $V_m$, cell membrane is destabilized and nanometer-sized pores are generated in the cell membrane as a result of pulse application. Pulses are applied repeatedly for a short duration (1 ms) with maintaining a fixed temporal gap between two pulses. It is assumed that the mass transfer takes place only when the pulse is off. The drug transport from extracellular to the intracellular domain occurs in the resealing period of the cell membrane. The maximum number of pores are created near the poles $\Psi=0$, $\pi$, in the setup as shown in Fig. \ref{electrode1}, as transmembrane potentials are maximum at those locations; whereas, no new pores are formed at $\Psi=\frac{\pi}{2}$, $\frac{3\pi}{2}$ as almost negligible transmembrane potential induced at those particular locations. 

%\section{Model Development and Mathematical Equations}

A uniform electric field and non-uniform electric potential has been developed on the arrangement of two parallel electrodes. The potential ($\phi$) distribution inside the domain is obtained by solving the  Laplace equation given as \cite{Mondal2021},
\begin{eqnarray}\label{6eq1}
	\nabla^2 \phi=0.
\end{eqnarray}
%with boundary conditions: $\phi(x=0)=\phi_0$ and $\phi(x=L)=\phi_L$, where $\sigma$ is the extracellular conductivity and considered that it is constant in this study.\\
The uniform electric field ($E$) throughout the domain $\Omega$ is obtained by taking the magnitude of potential gradient expressed as, 
\begin{eqnarray}\label{6E}
E= \lvert \vec{\nabla} \phi \lvert.
\end{eqnarray}
%In each cell membrane of the tissue the transmembrane potential  $V_m $, due to the effects of uniform electric field $E$, is developed  by DeBrurin and   Krassowka \cite{Krassowska1999} and defined as,
%\begin{equation}\label{transm_potential}
%V_m=1.5E\times r_c\cos \Psi ,
%\end{equation}
%where $r_c$ is the radius of the cell and  $\Psi$ is the angle between the direction of electric field  and  normal to the cell membrane at the position in which $V_m$ is calculated.\\ 

\subsection{Transmembrane potential and pore calculations}
The transmembrane potential $(V_m)$ was initially determined on a spherical cell in a uniform electrical field by Neumann et al. \cite{Neumann1989}. Later, DeBruin and Krassowska \cite{Krassowska1999} advanced their mathematical model to determine the pore density and its relation with the transmembrane potential. In the present study, we have used the model developed by DeBruin and Krassowska \cite{Krassowska1999} for the investigation of drug delivery into single-cell.

\begin{figure}[h!]
	\centering
	\begin{tikzpicture}[scale=0.4]
		
		%\draw[fill,blue!20] (0,0) circle (8cm);
		\draw[ultra thick] (0,0) circle (8cm);
		%	\draw[fill,black] (0,0) circle (0.08cm);
		%	\draw[ultra thick] (0,0) -- (6.9,4);
		%	\draw (3,1.8) node[anchor=north west]{\large $R$};
		\fill [black,even odd rule] (0,0) circle[radius=3cm] circle[radius=2.7cm];
		
		\draw[ultra thick] (0,0) -- (2.8,0);
		\draw[ultra thick] (0,0) -- (2.1,2);
		\draw[] (0.8,0) .. controls (1.0,0.4) .. (0.6,0.6);
		\draw (1.3,0.5) node[]{$\theta$};
		
		\draw[ultra thick,dashed, <->] (-2.75,0) -- (0.05,0);
		\draw (-1.2,.6) node[]{\large $a$};
		
		\draw[ultra thick,dashed, <->] (-8,0) -- (-2.95,0);
		\draw (-5.5,.6) node[]{\large $2a$};
		\draw (0,6) node[]{Extracellular space};
		\draw (0,-2) node[]{Cell};
		\draw[ultra thick,dashed, <-] (2.6,-1.5) -- (10,-3.5);
		\draw (11,-4) node[]{Cell membrane};
		\draw (3.5,0) node[]{\large $S$};
	\end{tikzpicture}
	\caption{Schematic diagram of a spherical cell with radius $a$ immersed in extracellular space of thickness $2a$.}
	\label{single_cell}	
\end{figure}
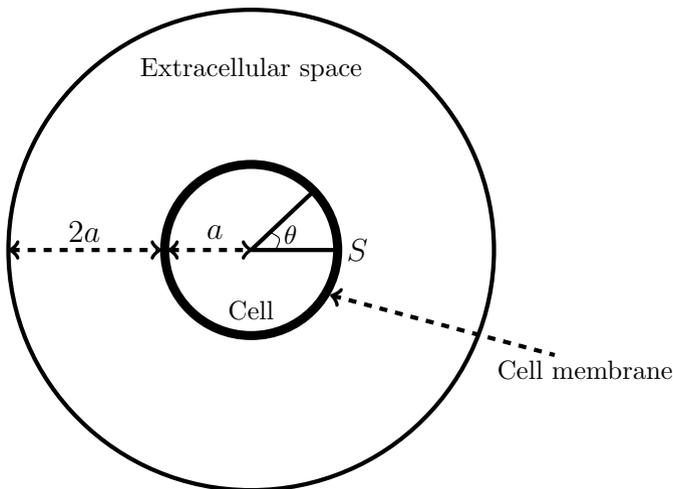
We consider a single-cell electroporation in which the spherical cell with radius $a$ is immersed in a spherical shell of extracellular space with radius $3a$. A uniform electric field is induced from the boundary of the extracellular space to destabilize the cell membrane.  The physical structure of the model comprising single-cell is schematically portrayed through Fig. \ref{single_cell}.

The transmembrane potential $V_m$ on the cell membrane due to the application of electric field is the  difference of potentials in intracellular and extracellular domains. As both the domains are source-free, the potentials are calculated using the Laplace equations as \cite{Krassowska1999},
\begin{align}
	\nabla^2\Phi_i&=0\quad \text{in intracellular space}, \label{6eq3}\\
	\nabla^2\Phi_e&=0\quad \text{in extracellular space},\label{6eq4}
\end{align}
where $\Phi_i$ and $\Phi_e$ are the intracellular and extracellular potentials. The uniform  external field $E$ is assumed at the outer boundary  and the $\Phi_e$ is obtained as,
\begin{equation}
	\Phi_e=-3aE\cos \theta.
	\label{6eq5}
\end{equation}
The current density across the cell membrane $(S)$ is given by
\begin{equation}
	-\hat{n}\cdot\left(\sigma_i\nabla\Phi_i\right)=-\hat{n}\cdot\left(\sigma_e\nabla\Phi_e\right)=C_m\frac{\partial V_m}{\partial t}+g_1\left(V_m-E_1\right)+N i_{ep}, 
	\label{6eq6}
\end{equation}
where $\hat{n}$ is the unit vector normal to the membrane's surface. $\sigma_i$ and $\sigma_e$ are the intracellular and extracellular conductivities, respectively. $C_m$ is the specific membrane capacitance and $V_m=\Phi_i-\Phi_e$ on $S$. $g_1$ denotes the specific membrane conductance, $t$ denots the time and $E_1$ the reversal potential of the ionic current. The current ($i_{ep}$ ) in a single pore is given as \cite{DeBruin1998},
\begin{equation}\label{6current}
	i_{ep}=\frac{\pi r_m^2\sigma v_m RT}{Fh} \frac{e^{v_m-1}}{\frac{w_0e^{(w_0-nv_m)}-nv_m}{w_0-nv_m}e^{v_m}-\frac{w_0e^{(w_0+nv_m)}+nv_m}{w_0+nv_m}},
\end{equation}
where $v_m=V_m\left(\frac{F}{RT}\right)$ is the non-dimensional transmembrane potential.

The pore density $N(t,\theta)$ given by Krassowka and Filev \cite{Krassowska2007} is as follows,
\begin{equation}\label{6eq8}
	\frac{dN}{dt}=\alpha A \left[1- \frac{N}{N_0}A^{-q}\right],
\end{equation}
where $A=\exp \left[\left( \frac{V_m}{V_{ep}} \right)^2\right]$, $t$ the time, $\alpha$ the pore creation rate coefficient, $V_m $ the  transmembrane potential,  $V_{ep}$ the characteristic voltage of electroporation, $N_0$ the equilibrium pore density for the membrane area at $V_m = 0$ and $q$ is an electroporation constant. The total number of pores  ($N_P$) in the area $\Delta_{\theta}$ of cell membrane after the application of electric pulse (duration is $t_{ep}$) is obtained as,
\begin{equation}\label{6eq9}
	N_P(\theta)=\oint\limits_{\Delta_{\theta}}N(t_{ep})\,d\theta.
\end{equation}
\subsection{Pore resealing}
The pore area decreases with time after electroporation due to the membrane resealing effect, which can be expressed as,
\begin{equation}\label{6eq10}
	A_P(\theta,t)=\pi R_P^2 \cdot N_P(\theta) \exp\left(-\frac{t}{\tau}\right),
\end{equation}
where $\tau$ is the resealing time, 
$\Delta_{\theta}$ is the local area at $\theta$ and $R_P$ is the average radius of the pores in the electroporated cell membrane.

The mass transfer coefficient is time dependent and depends on the pore density. The mathematical formulation of the mass transfer coefficient $(\mu)$ is given as follows,
\begin{equation}\label{6mut}
	\mu(\theta,t)=\frac{A_P(\theta,t)}{\Delta_\theta} P,
\end{equation}
where $P$ is the permeability of drug particles across the  cell membrane.

\subsection{Drug transport phenomenon in the tissue}
The drug concentrations in extracellular space and in the reversibly electroporated cell are obtained by the mass transport equations as,
\begin{align}
	&\frac{\partial C_E}{\partial t}=\nabla\cdot(D_E \nabla C_E), \label{6eq12}\\
	&\frac{\partial C_{I}}{\partial t}=\nabla\cdot(D_I  \nabla C_I),\label{eq:mass_transport}
\end{align}
with the initial and boundary conditions
%\begin{equation}\label{6ICS-BCS}
%\left.\begin{split}
	%&C_E(x, y, 0)=\begin{cases}
	%C_0,& x=0,\\
	%~0,&\text{otherwise},
	%\end{cases}\\
	%&C_{RE}(x, y, 0)=0,\\
	%&\frac{\partial C_E}{\partial\mathbf{n}}=0,\\
	%\end{split}\qquad\right\}
	%\end{equation}
	\begin{equation}\label{6ICS-BCS}
		C_E(x, y, 0)=\begin{cases}
			C_0,& x=0,\\
			~0,&\text{otherwise},
		\end{cases}
	\end{equation}
	\begin{equation}
		C_{RE}(x, y, 0)=0,
	\end{equation}
	\begin{equation}
		\frac{\partial C_E}{\partial\mathbf{n}}=0.
	\end{equation}
	%\begin{align*}
	%&\mu_E=\left(\frac{1-\varepsilon}{\varepsilon}\right)\mu (t_{ep}) \exp\left(-\frac{t}{\tau}\right),\\
	%\mu=\mu (t_{ep})\exp\left(-\frac{t}{\tau}\right),
	%\end{align*}
	Here, the subscripts $E$ and $I$ denote the variables from extracellular and intracellular spaces, respectively. $C$ is the drug concentration and $D$ drug diffusivity. $C_0$ denotes input initial drug concentration. $\mathbf{n}$ is an outward normal vector to the tissue surface.
	
	\subsection{Interface conditions}
	The cell membrane is selective permeable and the permeability depends on the drug properties. The improved permeability of cell membrane $(\mu)$ is incorporated in the drug transport as,
	\begin{equation}\label{eq:interface_conditions}
		D_E\nabla C_E=D_I\nabla C_I=\mu(\theta,t)(C_E-C_I){{\bf n}}.
	\end{equation} 
	\section{Method of solution}
	The mass transport Eqs. \eqref{eq:mass_transport}--\eqref{eq:interface_conditions} are solved using the permeable interface method (PIM) proposed by Yadav and Dalal \cite{fvhmm-p}. The PIM can be described briefly as follows. The set of equations are solved using the finite difference method. The domain is discretized using the Cartesian mesh, say the grid point is denoted by $(i,j)$, where $i$ is index in $x$-direction while $j$ is in the $y$-direction. The central-difference scheme used to discretize the Eq. \eqref{eq:mass_transport} is as,
	\begin{align}
		\left(\delta_x\left(D\delta_xC\right)\right)_{i,j} + \left(\delta_y\left(D\delta_yC\right)\right)_{i,j} =0,
	\end{align} 
	where
	\begin{equation*}
		\left(\delta_x\left(D\delta_xC\right)\right)_{i,j}=\left\{\frac{D_{i+1/2,j}C_{i+1,j}-(D_{i+1/2,j}+D_{i-1/2,j})C_{i,j}+D_{i-1/2,j}C_{i-1,j}}{(\delta x)^2}\right\}
	\end{equation*}
	and
	\begin{equation*}
		\left(\delta_y\left(D\delta_yC\right)\right)_{i,j}=\left\{\frac{D_{i,j+1/2}C_{i,j+1}-(D_{i,j+1/2}+D_{i,j-1/2})C_{i,j}+D_{i,j-1/2}C_{i,j-1}}{(\delta y)^2}\right\}.
	\end{equation*} 
	Here, the subscript $(i+1/2,j)$ denotes the position $(x_i+\delta x/2,y_j)$ with step-size $\delta x$ in $x$-direction. Similarly, other indices are defined.
	
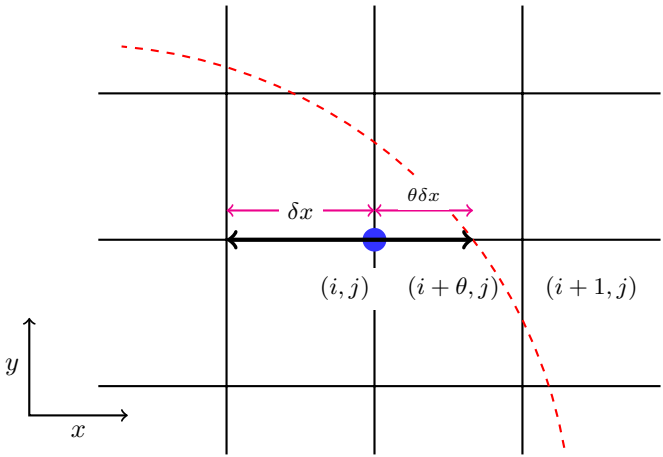
\begin{figure}[h!]
		\centering    
		\begin{tikzpicture}[scale=1.3]
			[inner sep=0pt,thick,dot/.style={fill=black,circle,minimum size=4pt},scale=0.8, every node/.style={scale=0.9}]
			\draw[step=1.5cm,thick] (0.2,0.8) grid (5.9,5.4);
			\fill[color=blue!80] (3,3)circle (0.8ex);
			
			\node[fill=white] at (3.8,2.5) {\small $(i+\theta,j)$};
			\draw [dashed,red,thick,domain=10:85] plot ({5*cos(\x)}, {5*sin(\x)});
			\node[fill=white] at (2.7,2.5) {\small $(i,j)$};
			%	\node[fill=white] at (1.2,2.5) {\small$(i-1,j)$};
			
			\node[fill=white] at (5.2,2.5) {\small $(i+1,j)$};
			%	\node[fill=white] at (3,5) {\small $(i,j+1)$};
			\draw[ultra thick,<->] (1.5,3)--(4,3);
			
			\draw[thick,magenta,<->] (3,3.3)--(4,3.3)node[midway,above,color=black,fill=white]{\footnotesize $\theta \delta x$};
			\draw[thick,magenta,<->] (1.5,3.3)--(3,3.3)node[midway,color=black,fill=white]{\small $ \delta x$};
			%	\draw[thick,magenta,<->] (1.8,3)--(1.8,4.5)node[midway,color=black,fill=white]{\small $ \delta y$};
			
			%	\draw[teal](4.78,1.5)node{\Large $\ast$}(4.5,2.1) node{\Large $\ast$} (4,3)node{\Large $\ast$} (3,4) node{\Large $\ast$} (2.15,4.5) node{\Large $\ast$} (1.5,4.75) node{\Large $\ast$} (-2.55,4.65) node{ {\Large $\ast$} \textcolor{black}{ $-$ Interfacial}};
			
			%	\fill[color=blue!80] (-3.5,4.05)circle (0.8ex) node[right]{ \textcolor{black}{~~$-$ Irregular}};
			
			%	\draw[thick] (1.5,1.5) circle (0.8ex) (3,1.5) circle (0.8ex) (1.5,3) circle (0.8ex)(4.5,4.5) circle (0.8ex) (-3.5,3.45) circle (0.8ex) node[right]{~~$-$ Regular};
			
			\draw[thick,<->] (-0.5,2.2)--(-0.5,1.2)node[midway,left]{$y$}--(0.5,1.2)node[midway,below]{$x$};
			%	\node at (2,2) {\Large $\omega_I$} (5,5) node {\Large $\omega_E$};
		\end{tikzpicture}
		\caption{Discretization near the interface.}
		\label{fig:PIM_computational_domain}
	\end{figure}

	However, on the grid points near the interface (as depicted in Fig. \ref{fig:PIM_computational_domain}), central-difference scheme can not be used directly. For this, the scheme used is as \cite{fvhmm},
	\begin{equation}\label{eq:modified_second_derivative}
		\left(\delta_x\left(D\delta_xC\right)\right)_i=\left\{D_{i+\theta/2}\frac{C_{i+\theta}^{-}-C_{i}}{x_{i+\theta}-x_i}-D_{i-1/2}\frac{C_{i}-C_{i-1}}{x_i-x_{i-1}}\right\}\Bigg/\left(\frac{x_{i+\theta}-x_{i-1}}{2}\right),
	\end{equation}  
	where $x_{i+\theta}=x_i+\theta \delta x$ for some $0<\theta<1$. $C_{i+\theta}^{-}$ is the limiting concentration at the point $x_{i+\theta}$ approaching from left side (Fig. \ref{fig:PIM_computational_domain}).
	
	The limiting concentrations at the interface are obtained using the linear interpolation from the left and right side, respectively, which also satisfy the interface conditions (Eq. \ref{eq:interface_conditions}) \cite{fvhmm}. The resulted system of equations is solved using the BiCGSTAB algorithm without preconditioning with maximum error between two consecutive iterative solutions falls below $10^{-15}$.

	\begin{table}[h!]
		\caption{The details of the parameters values used in the simulations.}
		\vspace{0.2cm}
		\label{6tab1}
		\centering
		\begin{tabular}{lllll}
			\hline
			Sym & Value  & Definition & Source   \\
			\hline 
			&&\\
			$r_c$ & 50 $\mu$m & Cell radius &\cite{Krassowska2007}  \\
			$\alpha$  & $10^9$ m$^{-2}$ s & Pore creation coefficient  &\cite{Krassowska2007}  \\
			$V_{ep}$ & 0.258 V& Characteristic voltage &\cite{Krassowska2007}  \\
			$N_0$ & $1.5\times10^9$ m$^{-2}$ & Equilibrium pore density   &\cite{Krassowska2007} \\
			$q$ & $2.46$ & Electroporation constant &\cite{Granot} \\
			$D_E$ & $10^{-3}$ mm$^{-2}$ s$^{-1}$ & Extracellular diffusion coefficient &\\
			$D_I$ & $10^{-4}$ mm$^{-2}$ s$^{-1}$ & Intracellular diffusion coefficient &\\
			$R_P$& $0.8$ nm & Pore radius &\cite{Granot} \\
			$P$ & $(0.1-1)$ mm s$^{-1}$ & Permeability of drug &\cite{Granot} \\
			$E$ & $(15-40)$ V mm$^{-1}$ & Electrical field & \\
			$C_0$ & 1 M & Initial drug concentration &\\
			$L$ & $3$ mm  & Edge length of the square & Fig. \ref{electrode1} \\
			$\phi_0$ & 25 V  & Potential at A & Fig. \ref{electrode1}\\
			$\phi_L$ & 0 V  & Potential at B & Fig. \ref{electrode1}\\
			$t_{ep}$ & 1 ms & Pulse length (ON TIME)& \\
			$t_M$  & 50 s, 100 s & Time for mass transfer (OFF TIME)& \\
			$PN$  & 20 & Pulse number& \\
			&&\\
			\hline
		\end{tabular}
		%\tablenote[t1n1]{This is an example of first tablenote entry. This is an example of first tablenote entry.}
		%\tablenote[t1n2]{This is an example of second tablenote entry.}
	\end{table}
	%Solving equation (\ref{2eq1}) with corresponding boundary conditions, the potential distribution in the domain $D_1$ is obtained as
	%\begin{equation}
	%\phi(x)=\frac{(\phi_L-\phi_0)}{L}x+\phi_0.
	%\end{equation}
	%Therefore, using equation (\ref{2E}) the uniform electric field in the tissue region $D_1$ is $E=\frac{(\phi_0-\phi_L)}{L}$.
	%Solving equation \eqref{pore_density}, from the equation \eqref{mut} the mass transfer coefficient is calculated as
	%\begin{equation}\label{soln_mut}
	%\mu(t)=\left(\frac{4\pi^2 R^{2}_Pr_c^2 P N_0A^q}{V_0}\right)\cdot \left[1- \exp \left(-\frac{\alpha t}{N_0 \cdot A^{q-1}}\right)   \right].
	%\end{equation}
	\section{Results and discussion}
	In this section, the effects of electroporation on drug transport in single-cell are analyzed through numerical experiments. The key parameters, such as drug permeability, electric field and pulse number are explored. In the numerical experiments, repeated pulses with a fixed voltage are applied to make the cell membrane permeabilized and retained this state for a longer period of time. Time gap between any two consecutive pulses is kept to be fixed (50 s or 100 s) for drug transport into the cell.  In order to incorporate the physiological situation, the parameters are taken from the literature as listed in the Table \ref{6tab1}. The results are obtained on a mesh size $250\times 250$ after ensuring grid independence outcomes.
	
	%\subsection{Pore density}
	%\begin{figure}[h!]
	%	\centering
	%	\includegraphics[scale=0.12]{2/pores}
	%	\caption{Pore density on the cell membrane, when the electrodes are fitted at the left and right sides. \textcolor{red}{The result is obtained for a pulse of 1 ms with $E=?$}}
	%	\label{fig:pores}
	%\end{figure}
	%If we consider the electrodes to be fixed on the left and right sides, the maximum pore density is obtained at $\theta=0,\pi$ (Fig. \ref{fig:pores}), which consequently improve the cell membrane permeability near these locations.

	\subsection{Validation}
	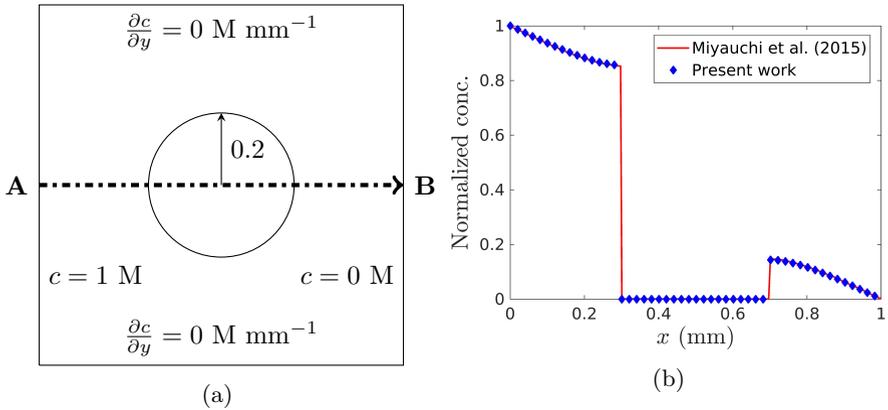
\begin{figure}[h!]
		\centering
		\begin{subfigure}{0.49 \textwidth}
			\centering
			\begin{tikzpicture}[scale=0.8]
				[inner sep=0pt,thick,dot/.style={fill=black,circle,minimum size=4pt},scale=0.8, every node/.style={scale=0.9}]
				[inner sep=0pt,thick,dot/.style={fill=black,circle,minimum size=4pt},scale=0.6, every node/.style={scale=0.6}]
				\draw {(0,0)}-- (6,0) node[midway,above] {$\frac{\partial c}{\partial y}=0$ M mm$^{-1}$} -- (6,3)node[midway,left]{$c=0$ M}--(6,6) -- (0,6) node[midway,below] {$\frac{\partial c}{\partial y}=0$ M mm$^{-1}$}--(0,3)-- (0,0)node[midway,right]{$c=1$ M};
				\draw (3,3) circle (1.2);
				\draw [ultra thick,dashdotted,->] (0,3)node[left]{\bf A} -- (6,3)node[right]{\bf B};
				\draw [->,>=stealth] (3,3) -- (3,4.2)node[midway,right]{0.2};
			\end{tikzpicture}
			\subcaption{}
		\end{subfigure}	
		\begin{subfigure}{0.49 \textwidth}
			\centering
			\includegraphics[width=\linewidth]{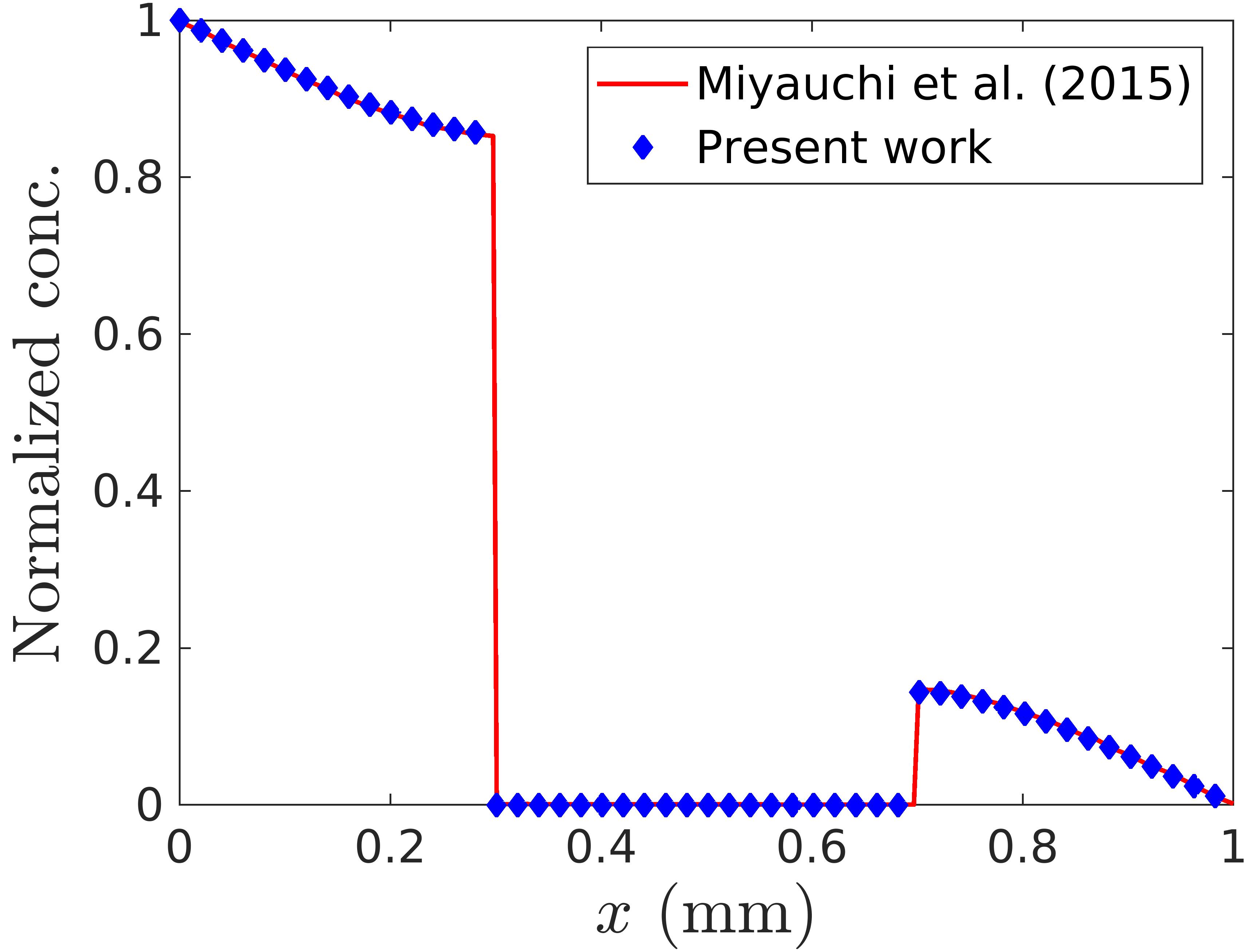}
			\subcaption{}
		\end{subfigure}	
		\caption{(a) Computational domain and (b) comparison of present results with the results of Miyauchi et al. \cite{immersed2015permeable_moving_membrane} along the line \textbf{AB} ($y=0.5$).}
		\label{fig:valiat}
	\end{figure}
	The results obtained from the model are validated with that of Miyauchi et al. \cite{immersed2015permeable_moving_membrane} and are shown in Fig. \ref{fig:valiat}. For this comparison, a square domain is considered with a cell placed at the center of it as shown in Fig. \ref{fig:valiat}a. $P=0$ is chosen, so no drug uptake is expected. The concentration distributions along the line \textbf{AB} are displayed in Fig. \ref{fig:valiat}b. It can be seen that the results are in very good agreement.
	\subsection{Drug distribution versus time}
	%\begin{figure}[h!]
	%	\begin{subfigure}{0.32 \textwidth}\centering
		%		\includegraphics[scale=0.13]{2/concet_contour_250_P_1_3_E150}
		%		\caption{$t=250$ s.}
		%	\end{subfigure}				
	%	\begin{subfigure}{0.33 \textwidth}\centering
		%		\includegraphics[scale=0.13]{2/concet_contour_500_P_1_3_E150}
		%		\caption{$t=500$ sec.}
		%	\end{subfigure}	
	%	\begin{subfigure}{0.33 \textwidth}\centering
		%		\includegraphics[scale=0.13]{2/concet_contour_1000_P_1_3_E150}
		%		\caption{$t=1000$ sec.}
		%	\end{subfigure}
	%\end{figure}
	
	\begin{figure}[h!]\centering
		\includegraphics[scale=0.22]{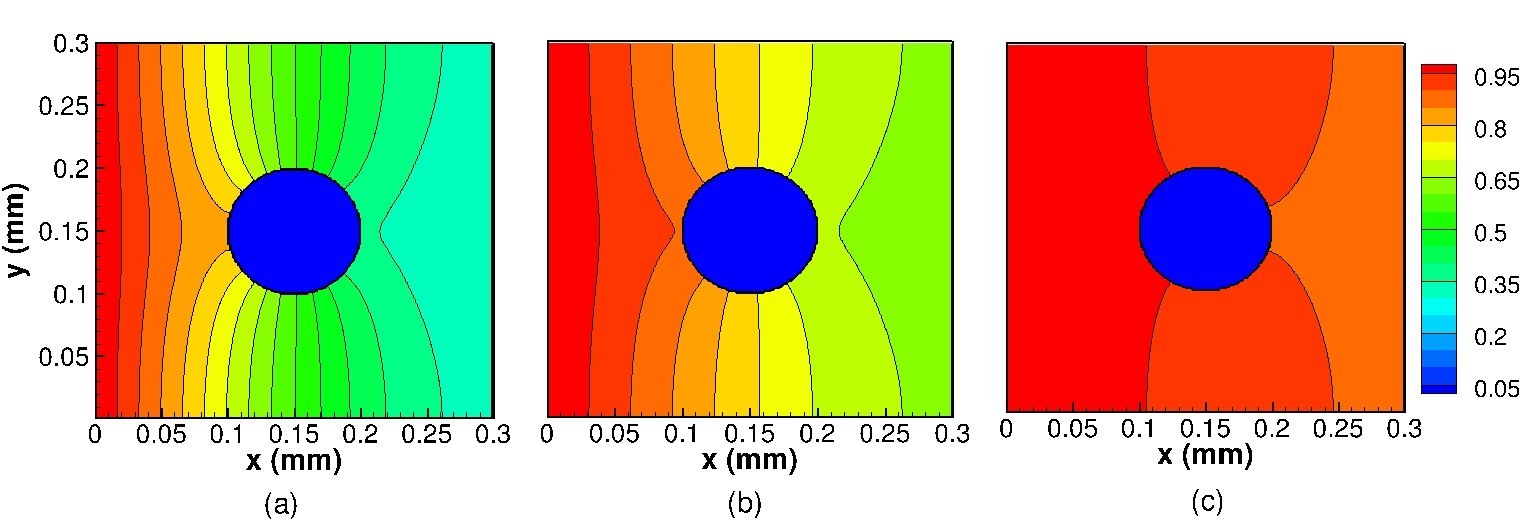}
		\caption{Contour plots of drug distributions at various times (a) $t=250$ s, (b) $t=500$ s and (c) $t=1000$ s for $P=0.1$ mm s$^{-1}$ and $E=15$ V mm$^{-1}$.}
		\label{E150_P3}	
	\end{figure}

	Fig. \ref{E150_P3} shows the contour plots of drug transport into the intracellular space from the extracellular region through the permeabilized cell membrane. It shows drug concentrations for different time durations, as $t=250$ s, 500 s, and for 1000 s. The drug uptake into the cell is very low due to the application of low voltage (15 V mm$^{-1}$) pulses. The reason is that when low voltage pulses are applied to the tissue, fewer number of pores are created in the cell membrane, and as a result, the membrane do not get sufficiently permeabilized for drug uptake.
	
	In order to improve membrane permeability, it is necessary to increase the electric field strength (i.e., $E>15$ V mm$^{-1}$) of the applied pulses in the experiments. Since a high voltage pulse enhances the number of pores and their area in the cell membrane (see Eq. \eqref{6eq8}), this gives rise to increase in mass transfer rate. The mass transfer rate may also increase for some drugs (basically small sized pharmaceutical molecules) that are highly permeable to the target cell membrane. This is due to the mass transfer coefficient $\mu$, which is directly proportional to drug permeability $P$ (Eq. \eqref{6mut}). So, several experiments are conducted to analyze the role of field strength of the applied pulses and drug permeability.  
	
	%Detailed discussion about the effects of significant parameters on cellular drug uptake are presented below.

	\subsection{Effects of electric field on drug penetration}
\begin{figure}[h!]
		\centering
		\includegraphics[scale=0.22]{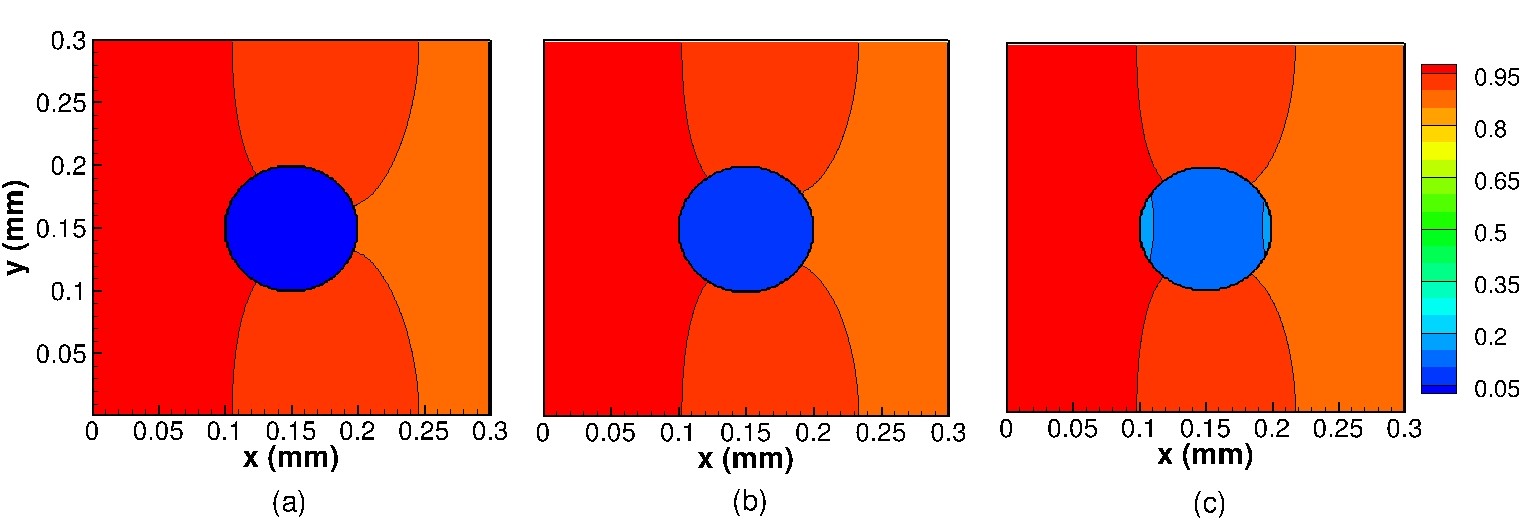}
		\caption{Effects of electric field (a) $E=15$ V mm$^{-1}$, (b) $E=25$ V mm$^{-1}$ and (c) $E=40$ V mm$^{-1}$ ($P=0.1$ mm s$^{-1}$, $t=1000$ s).}
		\label{fig:E}
	\end{figure}

	The experiments are conducted on the application of 20 pulses of 1 ms with three different electric fields as, $E=15$, 25 and 40 V mm$^{-1}$. Fig. \ref{fig:E} shows the effects of electric field on drug uptake. From Figs. \ref{fig:E}a, \ref{fig:E}b, it can be noticed that the drug uptake is very less due to insufficient permeabilization of the cell membrane with low voltage pulses ($E=15$, 25 V mm$^{-1}$). So, $E=15$, 25 V mm$^{-1}$ are not suitable voltages for introducing sufficient drug uptake into the cell. Another experiment for $E=40$ V mm$^{-1}$ with the same value of drug permeability ($P=0.1$ mm s$^{-1}$) is conducted to observe the effects of high electric pulse on mass transport into the cell. The results are shown in Fig. \ref{fig:E}c. A significant improvement in drug uptake is noticed; thus, sufficiently strong electric field is required in order to permeabilize the cell membrane appropriately. As indicated in Fig. \ref{fig:E}c, drugs enter the cell through both the sides ($\theta=0$, $\pi$), where maximum pores are created.
	\begin{figure}[h!]
		\centering
		\includegraphics[scale=0.22]{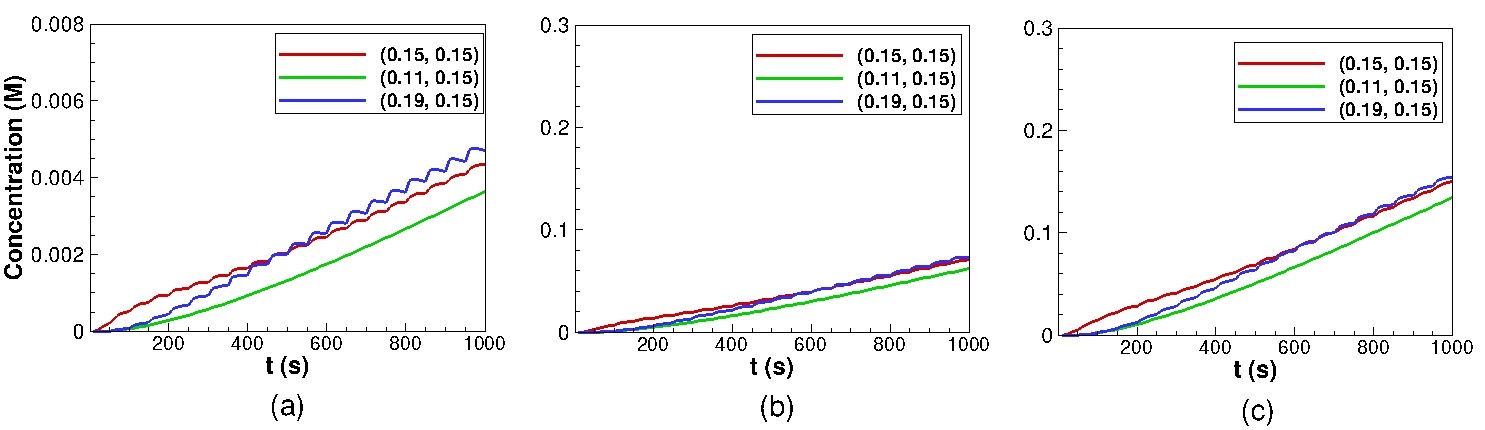}
		\caption{Concentration vs time inside the intracellular space for (a) $E=15$ V mm$^{-1}$, (b) $E=25$ V mm$^{-1}$ and (c) $E=40$ V mm$^{-1}$ ($P=0.1$ mm s$^{-1}$).}
		\label{fig:E_points}
	\end{figure}
	
	Fig. \ref{fig:E_points} shows the concentrations against the time obtained from some selected points inside the cell. The selected points are $(0.11,0.15)$, $(0.15, 0.15)$ and $(0.19,0.15)$. Clearly, one can observe that the drug concentration increases with time owing to the drug uptake. Effects of several pulses can be seen. On a given pulse, drug uptake initially improves and then it gets saturated due to the resealing effect, which subsequently requires another shot of pulse for faster uptake. Clearly, the intracellular concentration increases with increasing $E$. The maximum intracellular concentration is achieved for $E=40$ V mm$^{-1}$.

	%Drug concentration distributions along the line $y=0.15$ are also shown in the Fig. \ref{fig:E_cross-section}. The result is obtained at the end of the 20 pulses of 1 ms (one pulse of 50 sec). The intracellular drug concentration is much lower than the extracellular one. The drug concentration is higher at the end points of intracellular space while lower in its center owing to the uptake from both the sides (left and right, for $\theta=0, \pi$). It is expected that the intracellular drug concentration will eventually settle down to the steady state. 
	
	%\subsection{Drug distribution versus $x$}
	
	\begin{figure}[h!]
		\centering
		\includegraphics[scale=0.2]{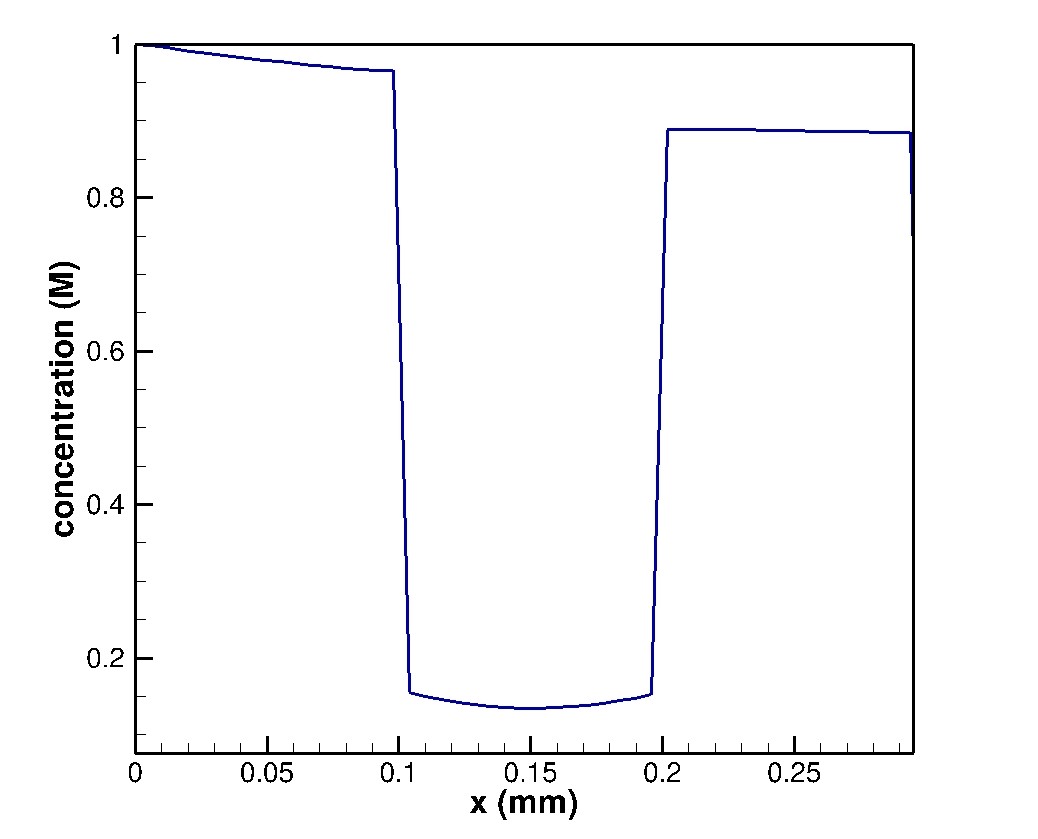}
		\caption{Concentration across a cross-section of domain at $y=0.15$ for $P=0.1$  mm s$^{-1}$ and $E=40$ V mm$^{-1}$.} 
		\label{cross-section_E400}
	\end{figure}
	Fig. \ref{cross-section_E400} displays the concentration distribution of drug along the line $y=0.15$ in the tissue domain.  The result is obtained at the end of the process in which 20 pulses of 1 ms with maintaining a 50 s temporal gap between the two consecutive pulses are considered. The figure shows that the extracellular concentration of the drug is much higher than the intracellular concentration. The use of a high electric field and the application of repeated pulses improve drug uptake into the cell. Therefore, our goal is to choose the electroporation parameters based on both the properties of a drug as well as the amount of drug that needs to be introduced into the cell.

	\subsection{Effects of drug permeability on drug penetration}
	\begin{figure}[h!]
		\centering
		\includegraphics[scale=0.22]{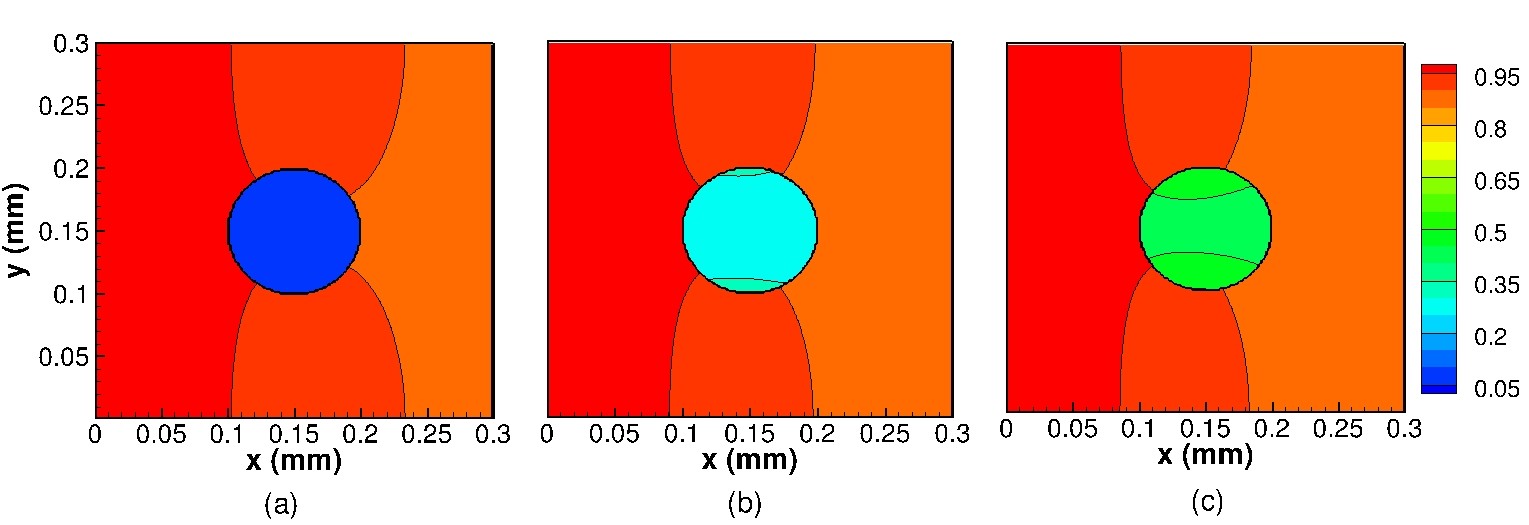}
		\caption{Effects of drug permeability (a) $P=0.1$ mm s$^{-1}$, (b) $P=0.5$ mm s$^{-1}$ and (c) $P=1$ mm s$^{-1}$ ($E=25$ V mm$^{-1}$, $t=1000$ s).}
		\label{fig:P}
	\end{figure}
	In order to observe the effects of drug permeability on cellular uptake, numerical experiments are conducted with different permeability values for the voltage $E=25$ V mm$^{-1}$. The results shown in Fig. \ref{fig:P} explain that the drug uptake increases with the increase in $P$, as the mass transfer rate increases with $P$. This shows that the drugs having higher permeability are required when low voltage fields are employed. It is evident from Figs. \ref{fig:P}b, c that a large amount of drug ($>0.2$ M) has entered the cell, which may be sufficient to treat the infected cell. Thus, the drugs with permeability greater than 0.005 can be chosen to inject it into the cell for $E=25$ V mm$^{-1}$.
	%This shows that the high permeable drug is required for low voltage, whereas, the low permeable drug can show better uptake with high voltage.
	%\textbf{The figure depicts that  drug uptake into the cell increases with the increase of $P$ for the voltage $E=250$ V mm$^{-1}$. So, we may choose this electric filed for sufficient cellular drug uptake. Fig. \ref{E400_P_change} represents the drug distribution profiles in the tissue at $t=1000$  sec for $E=400$ V mm$^{-1}$ and different values drug probabilities. The graphs show that higher amount of drug enters into the cell for higher values of $P$ as mass transfer rate increases  with $P$. Moreover, at $t=1000$ the diffusion of the drug into the extracellular space is the same, while its mass transfer from the extracellular to intracellular region differs.}

	\subsection{Optimal choice of $E$ and $P$ for enhanced drug uptake}
\begin{figure}[h!]
		\includegraphics[scale=0.22]{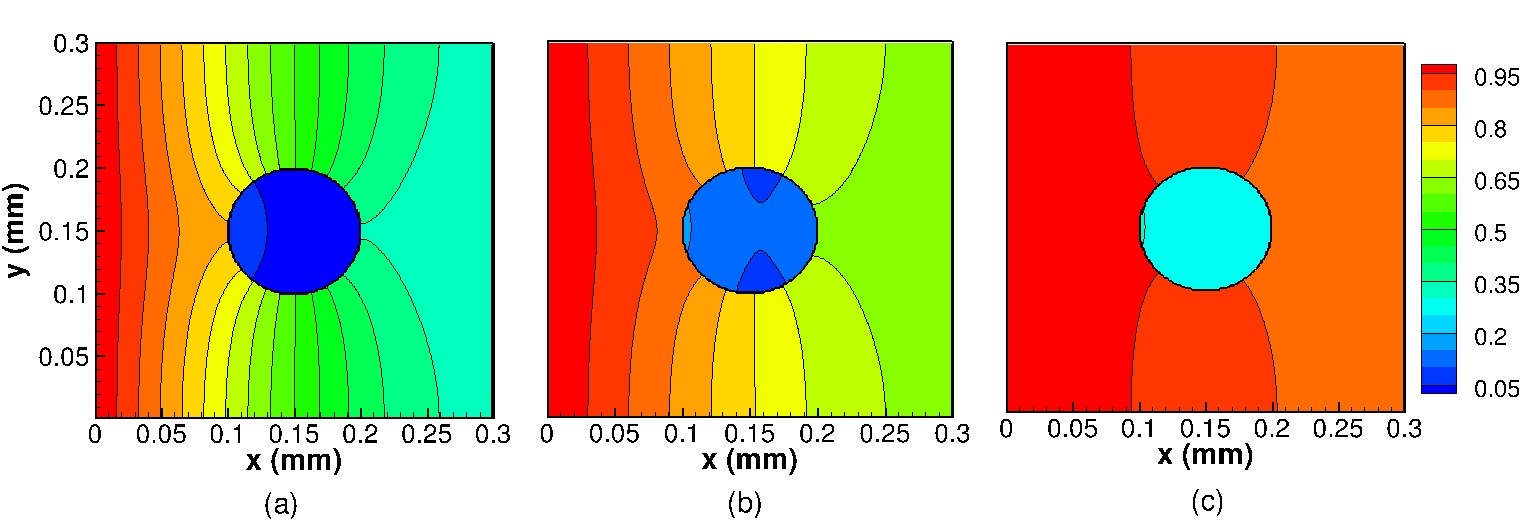}
		\caption{Contour plots of drug penetration at various times (a) $t=250$ s, (b) $t=500$ s and (c) $t=1000$ s for $P=0.5$ mm s$^{-1}$ and $E=25$ V mm$^{-1}$.}
		\label{E250_P_5_3}
	\end{figure}

	Fig. \ref{E250_P_5_3} shows the drug penetration into the cell at different times on application of pulses with a suitable voltage ($E=25$ V mm$^{-1}$) and appropriate choice of drug permeability $P=0.5$ mm s$^{-1}$. From Fig. \ref{E250_P_5_3}a, it can be noticed that the drug uptake has started at $250$ s from the left side of the cell when a sufficient amount of drug is diffused nearby the cell. 
	%This is due to the fact that a proper mass transfer rate occurs as increased values of both parameters electric field and drug permeability are considered. 
	On increasing time, the drug spreads out in the extracellular space and enters into the cell. 
	%Up to the end of the process ($t=1000$  s), the drug uptake into the cell increases continuously over time, that is observed by Figs. \ref{E250_P_5_3}b and \ref{E250_P_5_3}c. 
	Fig. \ref{E250_P_5_3}b illustrates that the drug enters into the cell through the pores at $\theta=\pi$ once some drug reach near this location due to the molecular diffusion. At the end of the process of drug delivery ($t=1000$ s), the drug concentration inside the cell is $\approx 0.25$ M by continuous drug uptake through both the left and right sides of the cell, which is shown in Fig. \ref{E250_P_5_3}c. The significant increase in drug concentration into the cell is obtained due to increased mass transfer rate on a proper choice of electric field and drug permeability.
	
	%\subsubsection{Comparison between two sets of parameter choice}
	\begin{figure}[h!]\centering
		\includegraphics[scale=0.22]{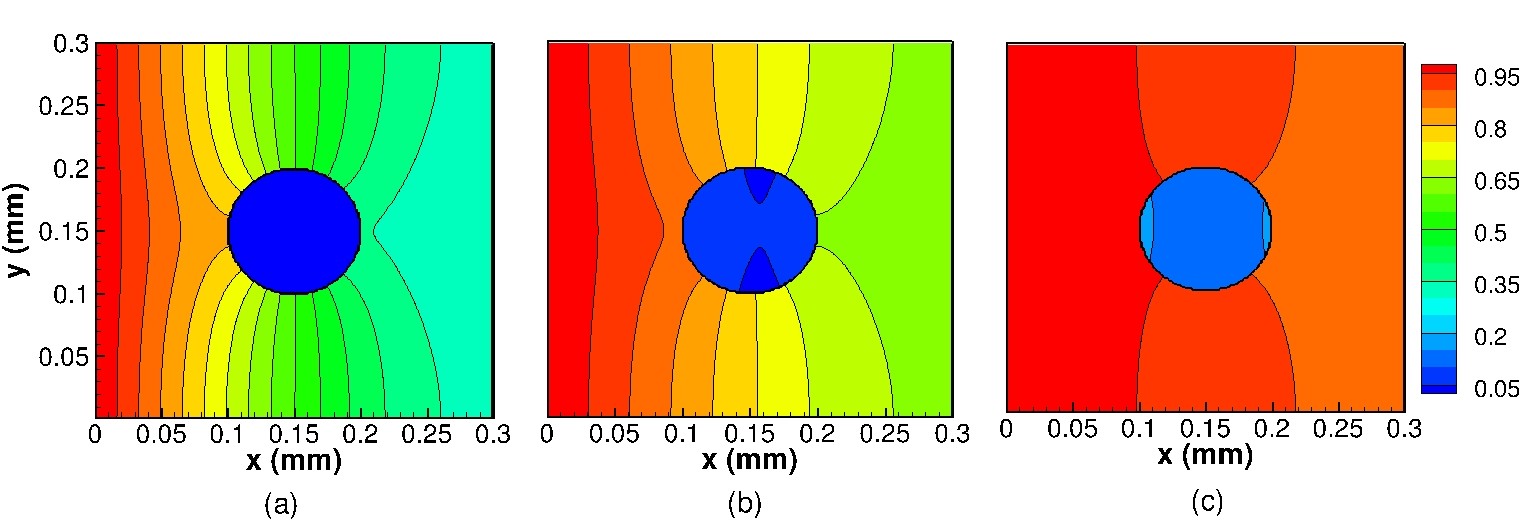}
		\caption{Contour plots of drug penetration at various times (a) $t=250$ s, (b) $t=500$ s and (c) $t=1000$ s for $P=0.1$ mm s$^{-1}$ and $E=40$ V mm$^{-1}$.}
		\label{E400_P_3}
	\end{figure}
	
	From the Fig. \ref{E400_P_3}, it can be seen that the cellular drug uptake (for $E=40$ V mm$^{-1}$ and $P=0.1$ mm s$^{-1}$) is almost equivalent to the drug uptake for $E=25$ V mm$^{-1}$ and $P=0.5$ mm s$^{-1}$ (see Fig. \ref{E250_P_5_3}). One can notice from Figs. \ref{E250_P_5_3}c and \ref{E400_P_3}c that the drug concentration into the cell is in the range 0.2 -- 0.3 M for both the cases. Therefore, these parameters' values ($E=20$ V mm$^{-1}$, $P=0.5$ mm s$^{-1}$ or $E=40$ V mm$^{-1}$, $P=0.1$ mm s$^{-1}$) can be taken into the consideration for the treatment of cancer cells by clinical experimentalists.
	
	%\begin{figure}[h!]\centering
	%	\includegraphics[scale=0.4]{2/concet_contour_P_1_2_E250_time}
	%	\caption{Drug distribution in the domain on time (a) $t=250$ sec, (b) $t=500$ sec and (c) $t=1000$ sec ($P=10^{-2}$ mm sec$^{-1}$ and $E=25$ V mm$^{-1}$).}
	%	\label{fig:time}
	%\end{figure}
	%Fig. \ref{fig:time} shows the contour plots of drug distribution for time durations, $t=250$, 500 and 1000 sec. Drug uptake takes place from the left side for time $t=250$ sec. (Fig. \ref{fig:time}(a)). On increasing time, the drug spreads out extracellularly outside the cell and enters into the cell from right side as well. The drug enters into the cell from two side only, not from the top and bottom sides. This is due to the high pore density in the cell membrane at $\theta=0,\pi$ generated on the application of electroporation on the left and right sides.
	\subsection{Effects of number of pulses}
\begin{figure}[h!]\centering
		\includegraphics[scale=0.2]{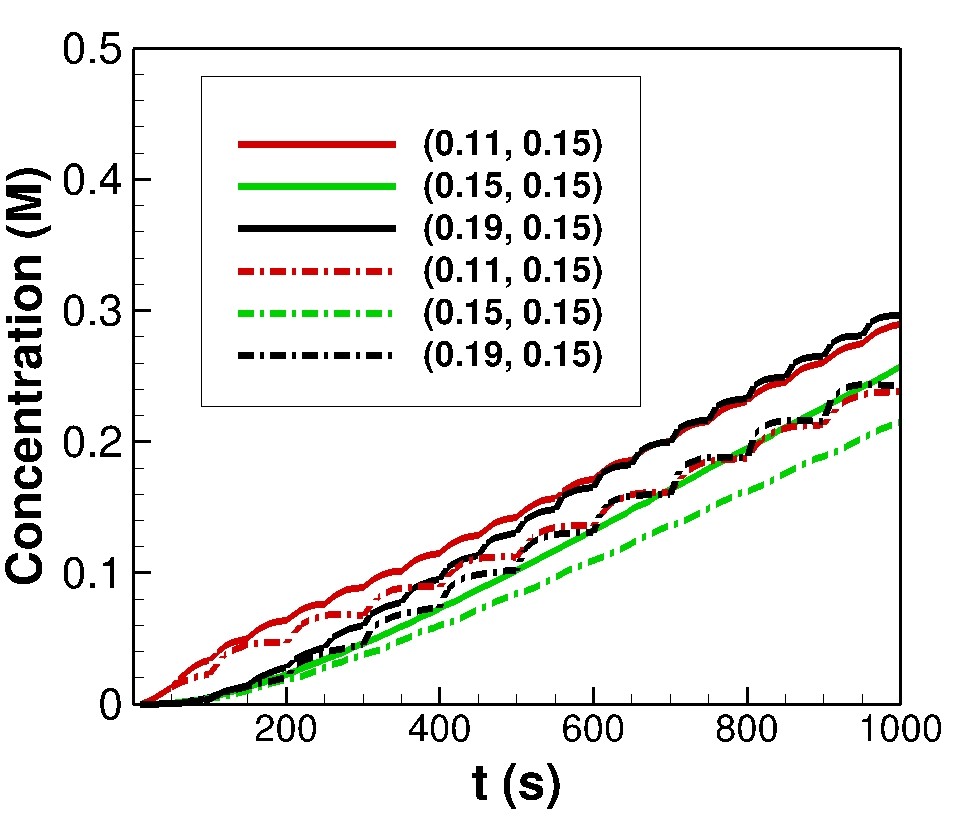}
		\caption{Effects of number of pulse shots on drug uptake ($E=25$ V mm$^{-1}$, $P=0.5$ mm s$^{-1}$). Solid lines `--' show for 50 s pulse gap while dashed lines `- -' represent 100 s pulse gap.}
		\label{fig:pulse_time}
	\end{figure}

	In order to study the effects of pulse shots on drug uptake, numerical experiments are performed in two ways; one, where a pulse is given after each 50 s while in other, a pulse is given after each 100 s. In the latter case, the time between two consecutive pulses is large, and consequently, for a given simulation time period the number of pulses is lesser. Concentration versus time on some intracellular points are plotted in Fig. \ref{fig:pulse_time}. The drug concentration increases over time for the application of pulses ($PN$: 20 if pulse gap is 50 s and 10 if pulse gap is 100 s). However, drug uptake improves on increasing the number of pulses. It is due to the resealing effects that will be pronounced if the time gap is more between two consecutive pulse shots.

	\subsection{Arrangement of electrodes}
\begin{figure}[h!]
		\centering
		\includegraphics[scale=0.22]{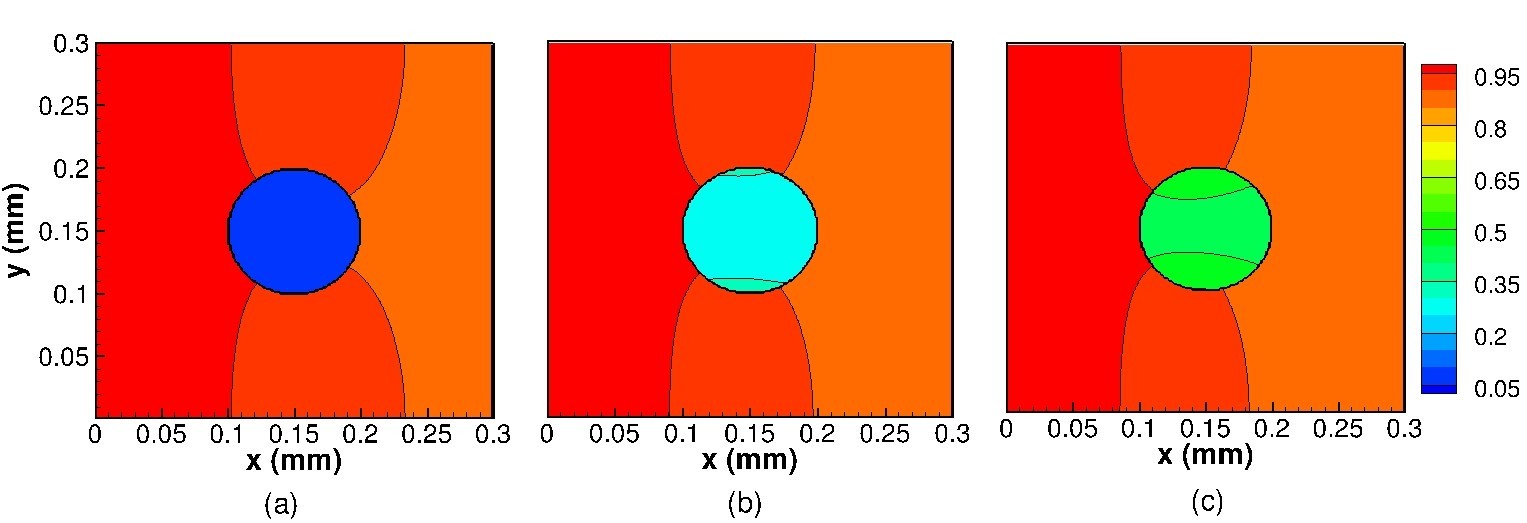}
		\caption{Drug distribution for the case where electrodes are placed at the top and bottom sides for (a) $P=0.1$ mm s$^{-1}$, (b) $P=0.5$ mm s$^{-1}$ and (c) $P=1$ mm s$^{-1}$ ($E=25$ V mm$^{-1}$).}
		\label{fig:P_rotate}
	\end{figure}

	In order to understand the effects of arrangement of electrodes, the experiments are conducted by setting electrodes on the top and bottom of the cell. In this case, the higher pore density is obtained at the top and bottom of the cell, i.e., for $\theta=\frac{\pi}{2}$ and $\frac{3\pi}{2}$. Fig. \ref{fig:P_rotate} shows drug distributions for different drug permeability values. In the extracellular region, drug distribution patterns are the same as obtained in earlier cases. However, drug uptake happens from the top and bottom parts of the cell, where pore density is high owing to the particular arrangement of electrodes.

	\section{Conclusions} 
	In this study, drug transport in a single-cell model has been studied. The effects of electroporation on drug uptake into the single-cell are explored. The transport across the cell membrane is incorporated using the permeable interface method available in the literature. Here, an advanced model in the area of electroporation drug delivery is analyzed as it provides the following important outcomes. 
	
	%A better generalized model is analyzed in comparison to the literature studies where the models are limited to the spherical coordinates.     
	
	Numerical experiments are conducted for electric fields $E= 15$, 25 and 40 V mm$^{-1}$ and for different drug permeabilities. It is noticed that the drug uptake into the cell is almost negligible on the application of significantly low voltage ($E=15$ V mm$^{-1}$) pulses whereas a significant improvement in drug uptake occurs for $E=40$ V mm$^{-1}$.
	%This is due to a fewer number of pores that have been generated with low voltage. As a result, the membrane does not get permeabilized  enough. For $E=40$ V mm$^{-1}$, a significant improvement in drug uptake is noticed. 
	A strong electric field is required to permeabilize the cell membrane properly, allowing a desired amount of drug to enter the cell. Based on the numerical experiments on the proposed model, it is learned that the suitable electric field is 25 V mm$^{-1}$  or more, and appropriate drugs whose permeability is at least $0.5$  mm s$^{-1}$ for getting a desired amount of drug into the targeted cell. This is a new finding from the prescribed model. Thus, in practice,  parameter values ($E=25$ V mm$^{-1}$, $P=0.5$ mm s$^{-1}$ or $E=40$ V mm$^{-1}$, $P=0.1$ mm s$^{-1}$) can be chosen to deliver pharmaceutical compounds into the targeted cell that needs  treatment.
	
	The drug uptake is initiated as soon as the application of pulse is completed. However, the drug uptake slows down due to the resealing effect, another shot of pulse is required to restore the mass transport. So, multiple pulses are required to increase the drug uptake and to achieve the desired level of drug absorption into the cell. The maximum drug uptake occurs through the poles at $\theta=0, \pi$ on setting electrodes on the left and right sides of the cell. Enhanced drug uptake is obtained for high permeable drugs, high voltage pulses, and by increasing the number of pulses.

%\backmatter
%
%\bmhead{Supplementary information}
%
%If your article has accompanying supplementary file/s please state so here. 
%
%Authors reporting data from electrophoretic gels and blots should supply the full unprocessed scans for key as part of their Supplementary information. This may be requested by the editorial team/s if it is missing.
%
%Please refer to Journal-level guidance for any specific requirements.

\bmhead{Acknowledgments}

The first-two authors are financially supported by Indian Institute of Technology Guwahati for conducting this work.

 \section*{Conflict of interest}

On behalf of all authors, the corresponding author states that there is no conflict of interest.

\bibliography{mybibfile}% common bib file


\section*{Statements \& Declarations}
\begin{itemize}
	\item The authors declare that no funds, grants, or other support were received during the preparation of this manuscript.
	\item The authors have no relevant financial or non-financial interests to disclose.
	\item The authors declare that the results/data/figures in this manuscript have not been published elsewhere, nor are they under consideration  by another publisher.
	\item The datasets generated during and/or analysed during the current study are available from the corresponding author on reasonable request.
	\item All authors contributed equally to this work.
\end{itemize}

\end{document}